\documentclass[10pt,conference]{IEEEtran}
\IEEEoverridecommandlockouts
\usepackage{cite}
\usepackage{comment}
\usepackage{amsmath,amssymb,amsfonts}
\usepackage{graphicx}
\usepackage{textcomp}
\usepackage{xcolor}
\usepackage{balance}
\def\BibTeX{{\rm B\kern-.05em{\sc i\kern-.025em b}\kern-.08em
    T\kern-.1667em\lower.7ex\hbox{E}\kern-.125emX}}
    
\usepackage[linesnumbered,ruled,vlined]{algorithm2e}

\SetKwInput{KwInput}{Parametrical input} 
\SetKwInput{KwData}{Measured input}
\SetKwInput{KwOutput}{Output}

\usepackage{graphicx}
\usepackage{caption}
\usepackage{subcaption}
\usepackage{tikz}
\usepackage{tikz-cd}

\begin{document}

\bstctlcite{IEEEexample:BSTcontrol}

\title{On Efficiently Partitioning a Topic in Apache Kafka\\
\thanks{This work was funded by the European Union’s Horizon 2020 research and innovation programme MARVEL under grant agreement No 957337. This publication reflects the authors views only. The European Commission is not responsible for any use that may be made of the information it contains.}
}

\author{\IEEEauthorblockN{Theofanis P. Raptis and Andrea Passarella}
\IEEEauthorblockA{\textit{Institute of Informatics and Telematics}, 
\textit{National Research Council}, 
Pisa, Italy. Email:
\{name.surname\}@iit.cnr.it}
\vspace{-0.9cm}
}

\maketitle

\begin{tikzpicture}[remember picture,overlay]
\node[anchor=south,yshift=10pt] at (current page.south) {\fbox{\parbox{\dimexpr\textwidth-\fboxsep-\fboxrule\relax}{
  \footnotesize{
    This work has been submitted to the IEEE for possible publication. Copyright may be transferred without notice, after which this version may no longer be accessible.
  }
}}};
\end{tikzpicture}

\begin{abstract}
Apache Kafka addresses the general problem of delivering extreme high volume event data to diverse consumers via a publish-subscribe messaging system. It uses partitions to scale a topic across many brokers for producers to write data in parallel, and also to facilitate parallel reading of consumers. Even though Apache Kafka provides some out of the box optimizations, it does not strictly define how each topic shall be efficiently distributed into partitions. The well-formulated fine-tuning that is needed in order to improve an Apache Kafka cluster performance is still an open research problem. In this paper, we first model the Apache Kafka topic partitioning process for a given topic. Then, given the set of brokers, constraints and application requirements on throughput, OS load, replication latency and unavailability, we formulate the optimization problem of finding how many partitions are needed and show that it is computationally intractable, being an integer program. Furthermore, we propose two simple, yet efficient heuristics to solve the problem: the first tries to minimize and the second to maximize the number of brokers used in the cluster. Finally, we evaluate its performance via large-scale simulations, considering as benchmarks some Apache Kafka cluster configuration recommendations provided by Microsoft and Confluent. We demonstrate that, unlike the recommendations, the proposed heuristics respect the hard constraints on replication latency and perform better w.r.t. unavailability time and OS load, using the system resources in a more prudent way.  
\end{abstract}


\section{Introduction}

Data stream processing is gaining more and more attention as a new implementation paradigm to manage real-time data-driven applications in numerous smart-* domains, such as cities \cite{9593258}, industry \cite{9039732}, networking \cite{s18082611}, and agriculture \cite{ANGELOPOULOS2020107039}. This paradigm is more suitable to modern application developers that consist of vertically separated engineering teams, each one managing a loosely coupled sub-system. In comparison to old-fashioned data-driven applications with centralized, shared data management systems, the distributed and non-synchronous nature of stream handling enables stakeholders to construct their sub-systems in an event-driven way which heavily relies to event-data passing \cite{10.1145/3448016.3457556}. The sub-systems get activated with local, asynchronous state updates in the presence of loosely coupled interactions which would typically need to be centrally managed and synchronized.

Apache Kafka was initially implemented so as to address the generic challenge of delivering extreme high volume event data to different consumers. Apache Kafka handles data as a partitioned write-ahead commit log on persistent storage and implements a pull-based messaging method to enable both real-time consumers such as online services and offline consumers such as Hadoop to consume those data at their own pace. During the last decade, Apache Kafka has become one of the most successful Apache open source products and is widely appreciated in the distributed systems community.

A representative case of Apache Kafka adoption is the largest professional social network on the Internet, LinkedIn. This network produces billions of events that reflect its over-300 million members’ social interactions and actions every day and routes them to its products. Product examples include “Who’s Viewed My Profile?” which is an info and news feeds, “People You May Know” which is an ML-based social and content recommendation system, and various other back-end monitoring and reporting systems like site health auditing processes as well as account fraud detection tools. Therefore, the ability to deliver high-volume event data in real time to both LinkedIn user-facing products as well as its back-end systems is highly important. With this objective under consideration, Apache Kafka has been developed as LinkedIn’s main online data handling framework \cite{10.14778/2824032.2824063}. 

Apache Kafka is a publish-subscribe data stream service, where a producer sends data to an Apache Kafka topic (a logical category of data) in a set of data brokers (the Apache Kafka cluster), and a consumer receives the data from the subscribed topic upon request. A topic may be stored in $\geq 1$ partitions (the physical storage of data in the Apache Kafka cluster), and all the partitions of the topic are distributed among the brokers of the cluster. Apache Kafka utilizes partitions to distribute a topic across many brokers for producers to send data in parallel, and also to enable parallel reception of consumers. Although Apache Kafka gives the capability to employ $4.000$ partitions per broker and $200.000$ partitions per cluster \cite{kafka-partitions}, it does not strictly define how each topic shall be efficiently divided to partitions or how shall the partitions be efficiently allocated to brokers. Moreover, to the best of our knowledge, there has been no research study in the state-of-the-art to address this emerging issue in a systematic manner, even in the case of a single topic. 

\subsection{Novelty of the Current Study}

As discuss in detail in section \ref{sec::related},  there has been no study in the state-of-the-art to  (i) model the Apache Kafka topic partitioning process in a fine-grained, combinatorial manner (even for a single topic), (ii) identify, formulate and characterize the emerging related optimization problem(s), and, (iii) provide efficient algorithmic solutions by taking into account the application requirements. Some questions arise, for example, how many partitions shall be allocated to a topic? How many brokers shall be used? Or, how to effectively respect the application constraints?

In the following sections we address such questions and provide an evaluation of our design. In section \ref{sec::basics}, we provide the Apache Kafka basics. In section \ref{sec::model}, we model the application constraints and requirements. In section \ref{sec::problem}, we formulate the problem that emerges from the modeling assumptions. In section \ref{sec::algorithms}, we design two heuristic algorithms for addressing the problem. In section \ref{sec::perf}, we conduct a performance evaluation against some configuration recommendations from the state-of-the-art and we highlight the strengths of our heuristics.


\subsection{Related works} \label{sec::related}

Although the related state-of-the-art has not researched the systematic formulation of topic partitioning in Apache Kafka deployments, there has been a notable corpus of works regarding various aspects of Apache Kafka functionalities. We briefly report those works here, so as the reader can grasp the current status of the area.

\begin{itemize}
    \item \emph{Apache Kafka fundamentals}. In \cite{10.1145/3459960.3459961}, the authors use formal methods to model the communication between producers and consumers in Apache Kafka. They also apply the model checking tools to verify five properties of the system. 
    In \cite{10.1145/3437120.3437283}, the authors develop a custom made Apache Kafka consumer, which enhances Apache Kafka with robust, scalable and smart filtering and queuing mechanisms to achieve effective rate adjustment. 
    In \cite{10.1145/3406601.3406612}, the authors promote the improvement of fault tolerance in the Apache Kafka pipeline architecture by defining optimal checkpoint interval values which impact the data recovery of the original Apache Kafka pipeline. The design is proved to reduce the number of lost data by using the optimal checkpoint interval time. 
    In \cite{8855525}, the authors analyze the structure and workflow of Apache Kafka and propose a queuing based packet flow model to predict performance metrics of Apache Kafka cloud services. 

    \item \emph{Apache Kafka applications.} In \cite{10.14778/2824032.2824063} the authors replicate Apache Kafka logs for various distributed data-driven systems at LinkedIn, including source-of-truth data storage and stream processing. In \cite{9016008}, the authors design a distributed cluster processing model based on Apache Kafka data queues, to optimize the inbound efficiency of seismic waveform data. In \cite{9378217}, the authors extend Apache Kafka by building an in-memory distributed complex event recognition engine built on top of Apache Kafka streams. In \cite{10.1007/978-3-030-84811-8_2}, the authors design a simulation platform enabling evaluations of future mobility scenarios, based on an Apache Kafka architecture. In \cite{10.1145/3148055.3148068}, the authors break the streaming pipeline into two distinct phases and evaluate percentile latencies for two different networks, namely 40GbE and InfiniBand EDR (100Gbps), to determine if a typical streaming application is network intensive enough to benefit from a faster interconnect. 
    In \cite{10.1145/3445945.3445949}, the authors propose a distributed framework for the application of stream processing on heterogeneous environmental data, which addresses the challenges of data heterogeneity from heterogeneous systems and offers real-time processing of huge environmental datasets through a publish/subscribe method via a unified data pipeline with the application of Apache Kafka for real-time analytics. In \cite{10.14778/3137765.3137771}, the authors find that filtering on large datasets is best done in a common upstream point instead of being pushed to, and repeated, in downstream components. To demonstrate the advantages of such an approach, they modify Apache Kafka to perform limited native data transformation and filtering, relieving the downstream Spark application from doing this. 
    
    \item \emph{Apache Kafka performance evaluation}. In \cite{9446179}, the authors propose an approach to track a specific vehicle over the video streams published by the collaborating traffic surveillance cameras. 
    In \cite{8258548}, the authors make an evaluation of several configurations and performance metrics of Apache Kafka in order to allow users avoid bottlenecks and eventually leverage some good practice for efficient stream processing. In \cite{9359224}, a performance study shows that estimates regarding the performance impact of different Apache Kafka configurations, based on experiences and perceptions, are not always true. 
    In \cite{10.1007/978-3-030-48340-1_58}, the authors propose an Apache Kafka-based load shedding engine which operates when the latency exceeded the threshold, and discards data in the Apache Kafka producer for quick overload control. 
    In \cite{10.1145/3491086.3492470}, the authors conduct an analysis of Apache Kafka’s network fault tolerance capabilities. Across different Apache Kafka configurations, they observe that Apache Kafka is fault-tolerant towards network faults to some degree, and they report observations of its shortcomings. 
    In \cite{10.1145/3093742.3093908}, the authors compare RabbitMQ with Apache Kafka based on the core functionalities of pub/sub systems and they venture into a qualitative and empirical comparison of the common features of the two systems. 

    \item \emph{Apache Kafka latency and reliability}. In \cite{8990298}, the authors test the impact of numerous configuration parameters on the reliability of Apache Kafka, including retry strategies and replications of partitions for fault tolerance. In \cite{9251089}, the authors build an Apache Kafka testbed using Docker containers to analyze the distribution of Apache Kafka's end-to-end latency. 
    In \cite{9153457}, the authors find that changing configuration parameters can significantly impact the guarantee of data delivery in Apache Kafka. 
    In \cite{8990246}, the authors introduce a tool for testing the reliability of Apache Kafka so as to study different data delivery semantics in Apache Kafka and compare their reliability under poor network quality. 

\end{itemize}

\section{Apache Kafka basics} \label{sec::basics}

\begin{figure}[t!]
\centering
        \includegraphics[width=\columnwidth]{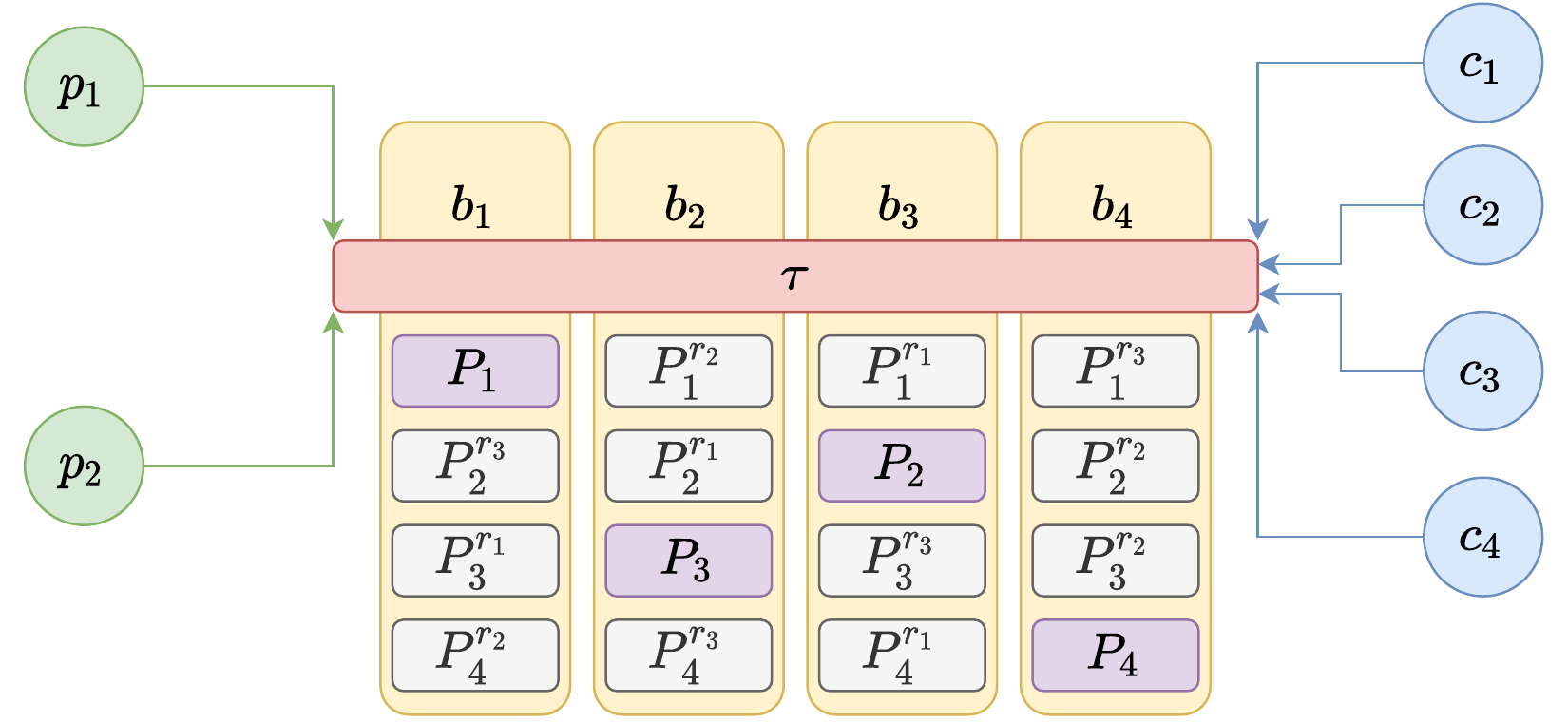}
        \caption{An example of an Apache Kafka configuration for a given topic $\tau$, with $p=2, c=4, b=4, P=4, r=3$. Leader partitions colored in purple, replicas in gray.}
        \label{fig::example}
        \vspace{-0.5cm}
\end{figure}

Apache Kafka provides a publish-subscribe data streaming service, in which $p$ producers send data to an Apache Kafka topic $\tau$ (a logical category of data) stored in a set of $b$ data brokers (the Apache Kafka cluster), and $c$ consumers read data from the topic. The topic is stored in $P \geq 1$ partitions (the physical storage of data in the Apache Kafka cluster), and all the partitions of the same topic are distributed among the brokers. Apache Kafka uses partitions to scale a topic across many brokers for producers to write data in parallel, and to further enable parallel reading of consumers. Every partition is able to be replicated across the Apache Kafka brokers for purposes of fault tolerance with a replication factor $r$. The replication helps in maintaining high availability in case one of the brokers goes down and is unavailable to serve the requests. A partition can be replicated across multiple replicas, each one stored on a different broker. One of the replicas is elected as leader and the remaining replicas are labeled as followers. Apache Kafka administers all the replicas automatically and ensures that they are kept in sync. The requests of both a producer and a consumer to a partition are served through leader replica. It is just the leader partition that manages all the data reads and writes of the interested producers and consumers, in a FIFO manner. An example of an Apache Kafka configuration is shown in Fig.~\ref{fig::example}.

When a consumer group is subscribed to a topic, each consumer instance in the group reads data from a different subset of the partitions in the topic in parallel. A consumer instance can read data from multiple partitions, while one partition has to be read by only one consumer instance within the same consumer group. Different consumer groups can independently consume the same data and no coordination among consumer groups is deemed necessary. Consequently, the number of partitions defines the maximum extent of parallelism within a consumer group.

Producers and consumers can  produce and consume increased volumes of data or generate requests at a very high rate and thus reserve broker resources, trigger network saturation and reduce the QoS of other consumers and brokers. Although Apache Kafka clusters can artificially apply quotas on requests in order to administer the broker resources utilized by producers and consumers, the quota functionality can not be applied in some application verticals that require high network bandwidth or high request rates.

The data topic partitioning reflects the extent of parallelism in Apache Kafka \cite{kafka-topicpartitions}. Latency significantly depends on the application design for handling data, and therefore optimizing it involves tuning the operational logic. On the other hand, the cluster throughput depends on the cluster parameters, and therefore, the broker and partition optimization can potentially increase efficiency. As the amounts of produced data increases, the number of partitions typically needs to scale up so as to prevent bottlenecks. High latency in the data pipeline can result in tremendous lag on the consumer side. The absence of an adequate number of partitions to achieve the required throughput can lead to time outs on the requests from producers to brokers and corresponding data loss. This major risk necessitates a balanced throughput and latency in the Apache Kafka cluster.

\section{Modelling application constraints and requirements} \label{sec::model}

\subsection{Data throughput}

In principle, as the number of partitions in an Apache Kafka cluster increases, so does also the throughput one can achieve. Therefore, a viable objective would be to maximize the number of partitions within the cluster:
\begin{equation}\label{eq::p}
\max P
\end{equation}
Writes to different partitions can be performed completely in parallel on both the producer and the broker side. As mentioned beforehands, in the consumer side, within a consumer group, Apache Kafka always maps the data of a single partition to a single consumer. Consumers can select both the topic and partition from which they want to read data. If consumers have not specified any partition, Apache Kafka can automatically choose more than one, based on the number of available partitions. A topic, is not able to support more consumers than partitions. Consequently, the extent of parallelism in the consumer side is bounded by the number of partitions able to be consumed:
\begin{equation}
P \geq c
\end{equation}
We denote the target Apache Kafka cluster throughput as $T$. If the throughput that can be achieved on a single partition for production and consumption ($T_p$ and $T_c$ respectively) is known (through measurements), then we need to have at least the following number of partitions:
\begin{equation}
P \geq \max \left(\frac{T}{T_p}, \frac{T}{T_c}\right)
\end{equation}
Except for throughput, there are a few other factors that have to be considered when optimizing the number of partitions. In some cases, having too many partitions may also have negative impact \cite{kafka-99perc}, which we highlight in the following sections.

\subsection{OS load}

Each partition corresponds to a directory in the broker's file system. In this directory, there are typically two files, one for the index and another for the data per log entry. Each broker maintains a single file handle for both the index and the data file of every log segment. Therefore, as the number of partitions increases, the open file handle limit in the OS must increase as well. Based on the above, if $H_{\max}$ denotes the maximum number of open file handles that a broker can tolerate, and $b$ the number of brokers, then the number of partitions in the Apache Kafka cluster is bounded by 
\begin{equation}
P \cdot r \leq b \cdot H_{\max}
\end{equation}

\subsection{Replication latency}

We consider as end-to-end latency in Apache Kafka the time required for the data to be published by the producer until the time that the data are read by the consumer. Apache Kafka provides access of data to a consumer after the point of time when the data has been replicated to all the in-sync replicas. Therefore, a significant portion of the end-to-end latency is considered the time to commit data. Typically, an Apache Kafka broker uses a single thread to replicate data from another broker for all partitions that they mutually share. Assuming a given replication factor $r$, we denote the replication latency as $l_r$. 

We can improve replication latency via building larger Apache Kafka clusters. For example, suppose that $x$ partition leaders reside on a broker and that the replication process necessitates in this case $y$ ms of time. If we add $z$ brokers in the same Apache Kafka cluster ($x > z$), each of the $z$ brokers needs to fetch only $x/z$ partitions on average from the initial broker. Therefore, the added latency of committing data will be in the order of just $y/z$ ms in this case, instead of $y$ ms. Depending on the application area, the application owner might enforce a requirement of a specific (end-to-end, and therefore) replication latency threshold, which we denote as $L$. Therefore, this threshold bounds the related system configurations as follows:
\begin{equation}
\frac{P\cdot r}{b} \cdot l_r \leq L
\end{equation}

\subsection{Unavailability}

The intra-cluster replication of Apache Kafka leads to higher availability and durability. However, when a broker fails, partitions with a leader on that broker become temporarily unavailable. Apache Kafka then automatically assign the leader role of those unavailable partitions to other replicas that continue serving the consumers. This is performed vis one of the Apache Kafka brokers designated as the controller, typically in serial fashion.

In specific cases, the observed unavailability can be proportional to the number of partitions. Indicatively, for the case of a broker that has a total of $x$ partitions, each with $y$ replicas: Roughly, this broker will be the leader for about $x/y$ partitions. If this broker fails, the $x/y$ partitions become unavailable at the same time. If $z$ ms time is required to elect a new leader for one partition, then it will require up to $z x/y$ ms to elect a new leader for all $x/y$ partitions. Therefore, for several partitions, the observed unavailability can be $z x/y$ ms plus the time taken to detect the failure. Therefore, given an observed unavailability time $u$ during a broker failure, the application unavailability requirement threshold $U$ shall bound the configuration as follows:
\begin{equation}\label{eq::u}
\frac{P}{b} \cdot u \leq U 
\end{equation}

\section{The problem} \label{sec::problem}

Based on the descriptions of section \ref{sec::model}, there are contrasting forces defining the number of partitions $P$ to be used, so choosing the ``right'' number of partitions per topic requires knowledge of many factors. Increasing the partition density (the number of partitions per broker) adds an overhead related to metadata operations and per partition request/response between the partition leader and its followers. Increasing the number of partitions in a cluster though will lead to increased parallelism of data consumption, which in turn improves the throughput of an Apache Kafka cluster; however, the time required to replicate data across replica sets will also increase \cite{granulate}.

Even though Apache Kafka provides some out of the box optimizations, the well-formulated fine-tuning that is needed in order to improve cluster performance is still an open research problem. Based on equations/inequalities \ref{eq::p}-\ref{eq::u}, we formulate the problem as the following program (the program variables are marked in bold font): 

\noindent \underline{Objective:}
\begin{align}
\max \quad
    & \pmb{P} \label{obj::P}
\end{align}
\noindent \underline{Constraints:}
\begin{align}
    \pmb{P} - \max \left(\frac{T}{T_p}, \frac{T}{T_c}, c\right) & \geq 0 && (\text{throughput}) \label{con::thru} \\
    \pmb{P} \cdot r - \pmb{b} \cdot H_{\max} &\leq 0 && (\text{OS load}) \label{con::os}\\
    \pmb{P}\cdot r \cdot l_r - \pmb{b} \cdot L &\leq 0 && (\text{replication latency}) \label{con::lat}\\
   \pmb{P} \cdot u - \pmb{b} \cdot U &\leq 0 && (\text{unavailability}) \label{con::avail}\\
   r \leq \pmb{b} &\leq B \label{con::bro}\\
    \pmb{P}, \pmb{b} &\in \mathbb{Z}^+ \label{con::z}
\end{align}

The objective function (\ref{obj::P}) maximizes the number of partitions in the cluster. Constraint (\ref{con::thru}) guarantees that the selected number of partitions renders the resulting throughput viable. Constraint (\ref{con::os}) guarantees that the OS load in terms of open file handles can be supported by the system in place. Constraint (\ref{con::lat}) guarantees that the replication latency does not exceed the maximum latency threshold provided by the operator. Constraint (\ref{con::avail}) guarantees that potential broker unavailability do not exceed the maximum unavailability threshold provided by the operator. Constraint (\ref{con::bro}) guarantees that the number of selected brokers does not exceed the actual number $B$ of available brokers in the cluster.

The problem is computationally intractable, since it is formulated as an integer program (all of the variables are restricted to be nonzero positive integers, through constraints (\ref{con::z})). This means that we are not able to optimally maximize the exact number of partitions needed to satisfy all the system constraints for any given instance of the problem in polynomial time. 

\section{Maximizing the number of partitions} \label{sec::algorithms}

\begin{algorithm}[!t]
\DontPrintSemicolon
  \KwInput{$T, c, r,  L, U, B$}
  \KwData{$T_p, T_c, H_{\max}, l_r, u$}
  \KwOutput{$P, b$}
    \For{$b = r;\ b \leq B;\ b++$}{ \label{line::brok}
        \For{$P =\left\lfloor\frac{b \cdot H_{\max}}{r}\right\rfloor;\ P \geq \max \left(\frac{T}{T_p}, \frac{T}{T_c}, c\right);\ P--$}{ \label{line::part}
            \If{$P \cdot r \cdot l_r \leq b \cdot L$ \textbf{and} $P \cdot u \leq b \cdot U$}{ \label{line::lat}
                return $P,b$; \label{line::sol}
            }
        }
    }
    return \emph{``No feasible solution found.''}\; \label{line::no}
\caption{BroMin} \label{alg::bromin}
\end{algorithm}

\begin{algorithm}[!t]
\DontPrintSemicolon
  \KwInput{$T, c, r,  L, U, B$}
  \KwData{$T_p, T_c, H_{\max}, l_r, u$}
  \KwOutput{$P, b$}
    \For{$b = B;\ b \geq r;\ b--$}{
        \For{$P =\left\lfloor\frac{b \cdot H_{\max}}{r}\right\rfloor;\ P \geq \max \left(\frac{T}{T_p}, \frac{T}{T_c}, c\right);\ P--$}{
            \If{$P \cdot r \cdot l_r \leq b \cdot L$ \textbf{and} $P \cdot u \leq b \cdot U$}{
                return $P,b$;
            }
        }
    }
    return \emph{``No feasible solution found.''}
\caption{BroMax} \label{alg::bromax}
\end{algorithm}

The naive way to solve the problem is to simply remove the constraint that $P$ and $b$ are integers (therefore enforce $b,P \in \mathbb{R}$ instead of constraints (\ref{con::z})), solve the corresponding linear relaxation of the integer program, and then round the entries of the solution to the linear relaxation. But, not only may this solution not be optimal, it may not even be feasible, as it could potentially violate some constraint. 

In order to address this challenge, we designed Algorithms \ref{alg::bromin} and \ref{alg::bromax} for a topic partitioning with ensuring adequate throughput, low OS load, low replication latency and high availability. Each of the two algorithms targets at fully exploiting a given broker's resources and at minimizing (Algorithm \ref{alg::bromin} - BroMin) or maximizing (Algorithm \ref{alg::bromax} - BroMax) the number of brokers used. Although the two approaches seem to follow an opposite line of though, they both try to maximize the number of partitions needed in the Apache Kafka cluster to satisfy the constraints, but with a different side-objective: use as few HW resources as possible in the first case and take advantage of all the available HW resources in the second case.

More specifically, both algorithms receive a set of parametrical inputs ($T, c, r, H_{\max},  L, U, B$) and a set of measured inputs ($T_p, T_c, l_r, u$). Each algorithm starts gradually increasing or decreasing correspondingly the number of brokers to be investigated, with a lowest number equal to the replication factor $r$ and highest equal to the number of available brokers $B$ (line \ref{line::brok}). Then, for the selected broker number, both algorithms start decreasing the number of investigated partitions $P$, starting from the maximum allowed number per broker w.r.t. open file handles, and finishing at the minimum allowed number w.r.t. the desired cluster throughput and consumer support (line \ref{line::part}). The algorithms then return as a solution, the first occurence of $P,b$ combination that satisfies the replication latency and unavailability time constraints (line \ref{line::lat}-\ref{line::sol}). As usual in the case of heuristic approaches, the inherent disadvantage is that if the algorithm fails to find a solution (line \ref{line::no}), it cannot be determined whether it is because there is no feasible solution or whether the algorithm simply was unable to find one. Further, it is usually impossible to quantify how close to optimal a solution returned by these methods are. In order to address the last concern, we perform a large scale simulation study in section \ref{sec::perf}.

\section{Performance evaluation} \label{sec::perf}

In this section we perform an extensive simulation evaluation of our algorithms in Octave 6.2.0. 

\subsection{Parameters' configuration}

We assign values to the algorithm measured inputs according to verified state-of-the-art measurements. For example, the per-partition throughput $T_p$ that one can achieve on the producer depends on configurations such as the batching size, compression codec, type of acknowledgement, replication factor, etc. However, as shown in \cite{kafka-benchmarking}, in general, one can produce at $T_p = 10$ MB/sec on just a single partition. The consumer throughput $T_c$ is often application dependent since it corresponds to how fast the consumer logic can process data. In our case we set it indicatively to $T_c = 20$ MB/sec. Furthermore, in \cite{kafka-topicpartitions}, it is noted that it can take around 5 ms to elect a new leader for a single partition, and therefore set the equivalent unavailability time to $u = 5$ ms. The same report, also claims that production Apache Kafka clusters can run with $ > 30.000$ open file handles per broker. In our case, in order to be able to capture also weaker settings, we set $H_{\max} = 10.000$. Next, from the report's contents, it can be assumed that the replication latency for a single partition (without considering other partitions) could be as low as $l_r = 1$ ms. Last but not least, the report states that leader election for a single partition can take $5$ ms, and therefore we set the unavailability time as $u = 5$ ms. The parametrical inputs $T, c, r, L, U$ and $B$, depend on the application requirements. Different applications might come with completely different values of those parameters. In our simulations, we variate the values of $c, B$ and $r$, and we set $T = 100$ MB/s, $L = 200$ ms and $U = 2000$ ms.

\begin{figure}[t!]
\centering
    \begin{subfigure}[b]{0.49\columnwidth}
    \centering
        \includegraphics[width=\columnwidth]{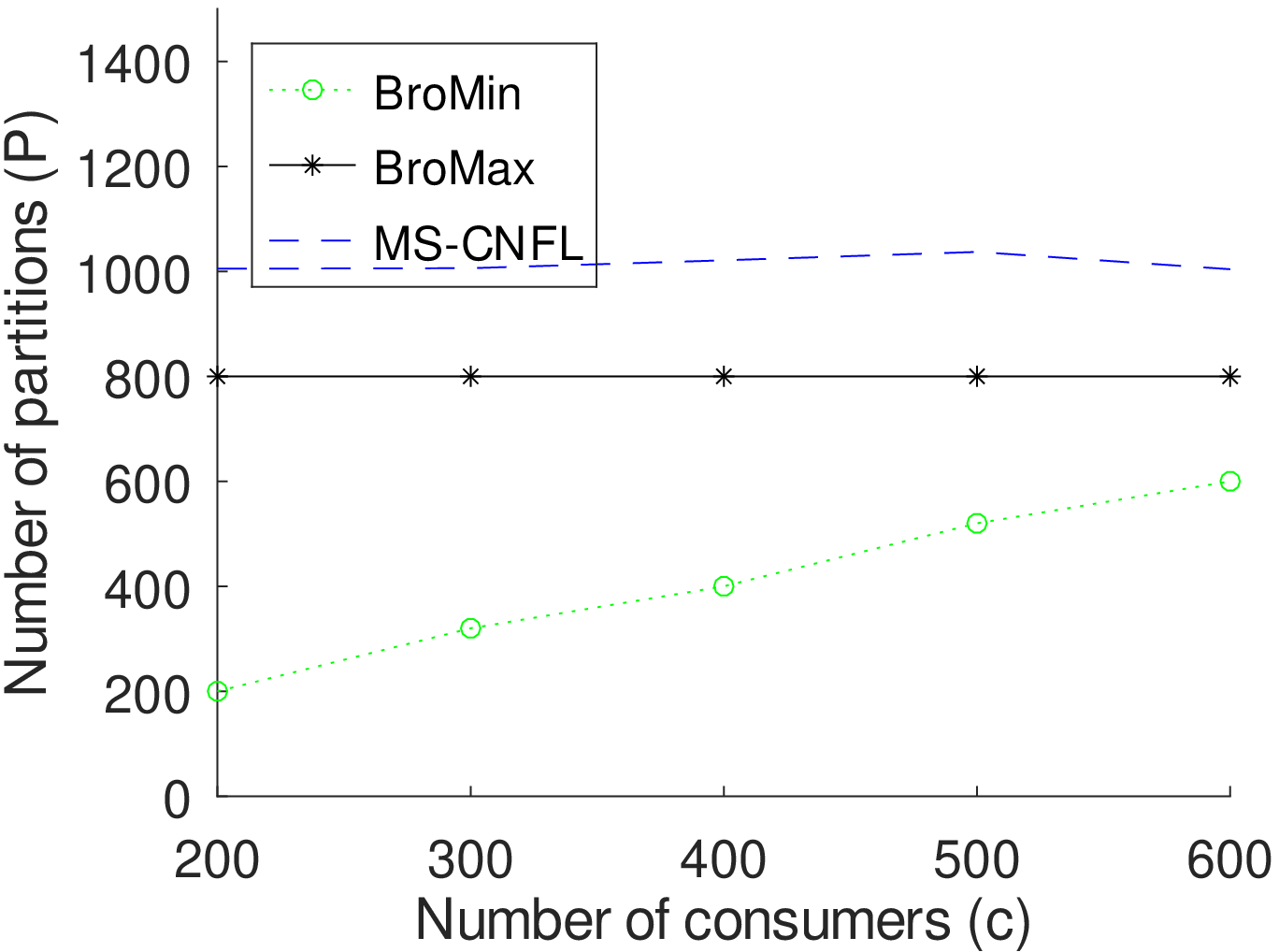}
        \caption{Partitions.}
        \label{fig::multiconsim-part}
    \end{subfigure}
    \begin{subfigure}[b]{0.49\columnwidth}
    \centering
        \includegraphics[width=\columnwidth]{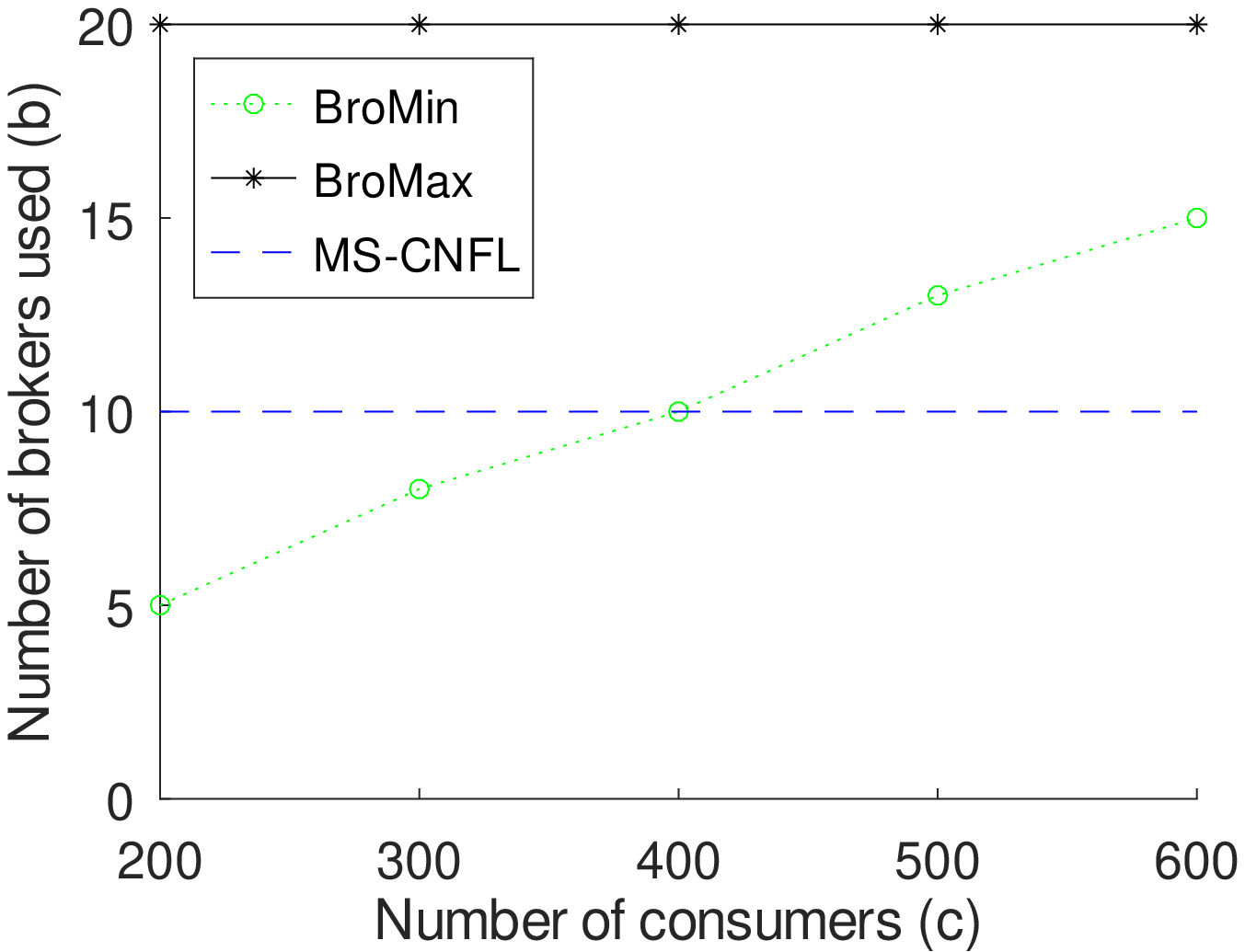}
        \caption{Brokers.}
        \label{fig::multiconsim-brok}
    \end{subfigure}    
    
    \begin{subfigure}[b]{0.49\columnwidth}
    \centering
        \includegraphics[width=\columnwidth]{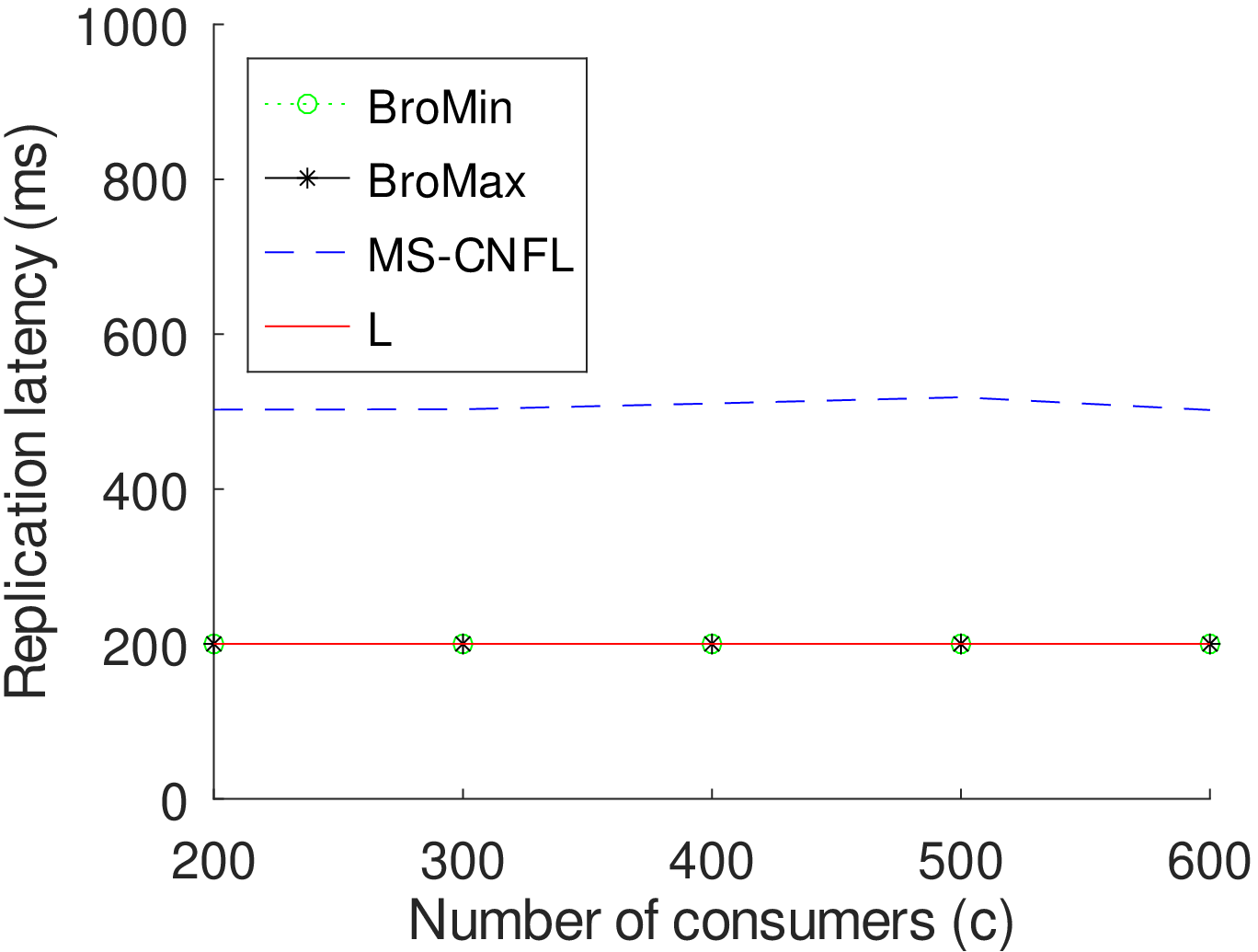}
        \caption{Latency.}
        \label{fig::multiconsim-lat}
    \end{subfigure}
    \begin{subfigure}[b]{0.49\columnwidth}
        \centering
        \includegraphics[width=\columnwidth]{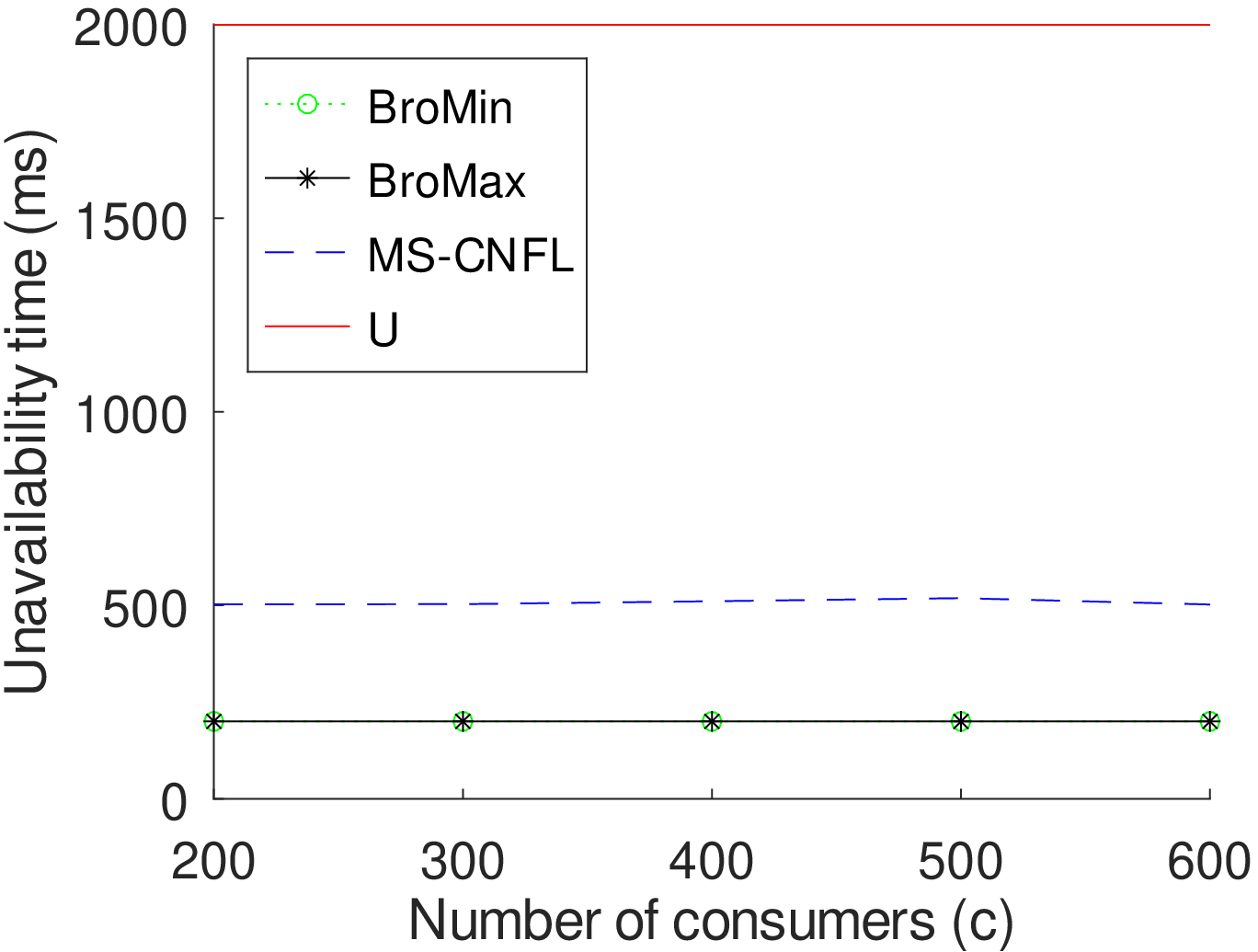}
        \caption{Unavailability.}
        \label{fig::multiconsim-unav}
    \end{subfigure}
    \begin{subfigure}[b]{0.49\columnwidth}
        \centering
        \includegraphics[width=\columnwidth]{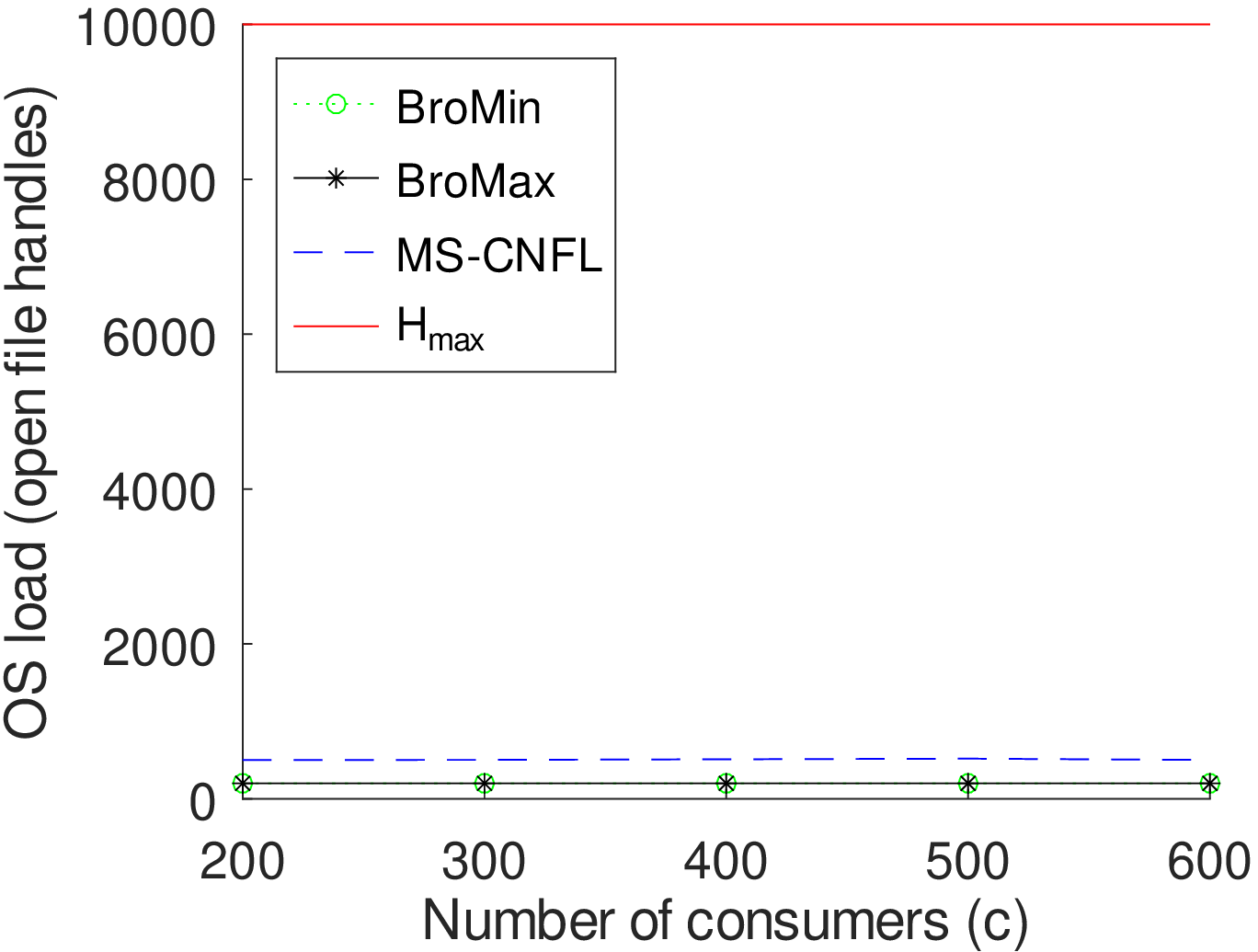}
        \caption{OS load.}
        \label{fig::multiconsim-load}
    \end{subfigure}
    \caption{Performance for variable numbers of consumers.}\label{fig::multiconsim}
    \vspace{-0.5cm}
\end{figure}

\begin{figure}[t!]
\centering
    \begin{subfigure}[b]{0.49\columnwidth}
    \centering
        \includegraphics[width=\columnwidth]{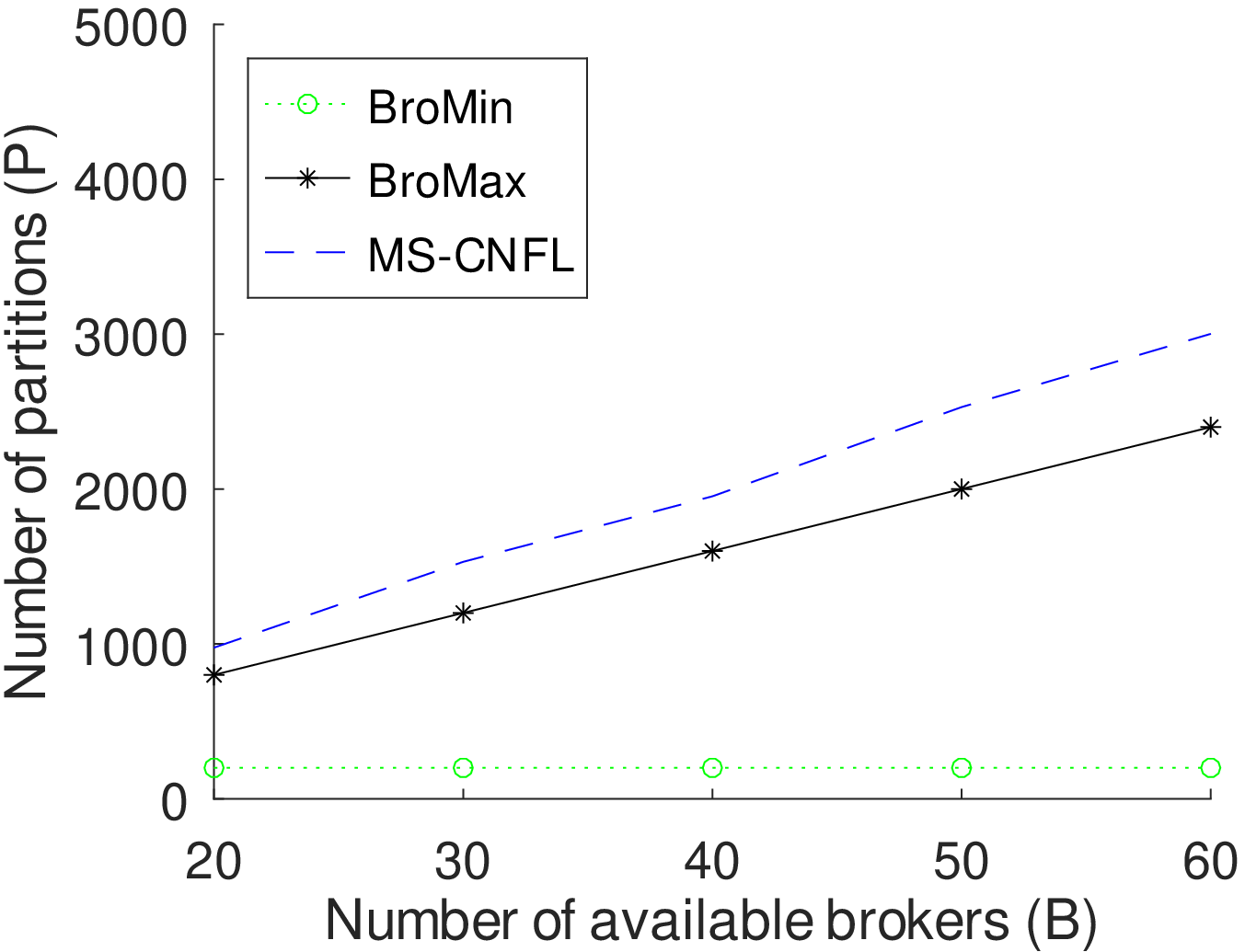}
        \caption{Partitions.}
        \label{fig::multibrosim-part}
    \end{subfigure}
    \begin{subfigure}[b]{0.49\columnwidth}
    \centering
        \includegraphics[width=\columnwidth]{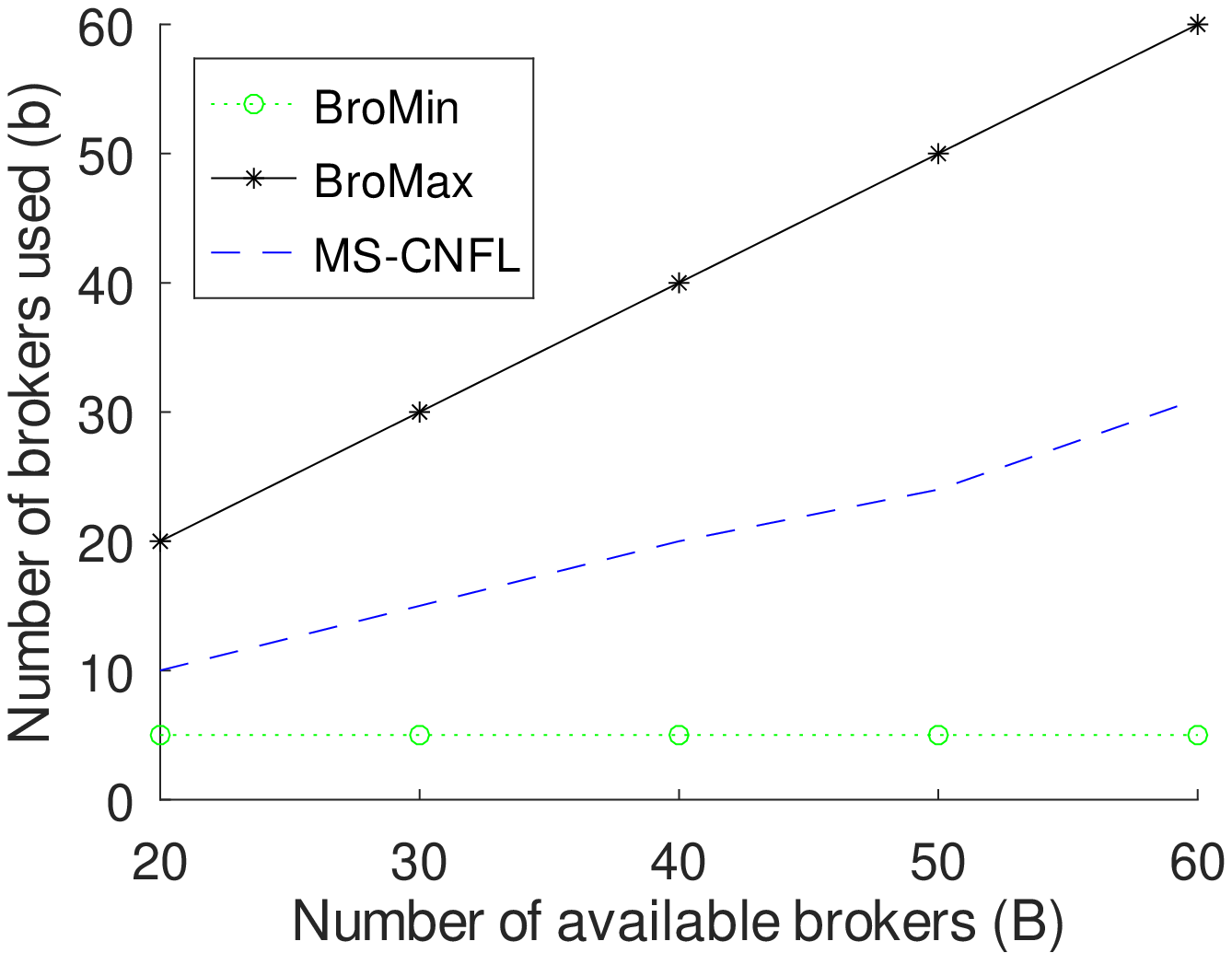}
        \caption{Brokers.}
        \label{fig::multibrosim-brok}
    \end{subfigure}    
    \begin{subfigure}[b]{0.49\columnwidth}
    \centering
        \includegraphics[width=\columnwidth]{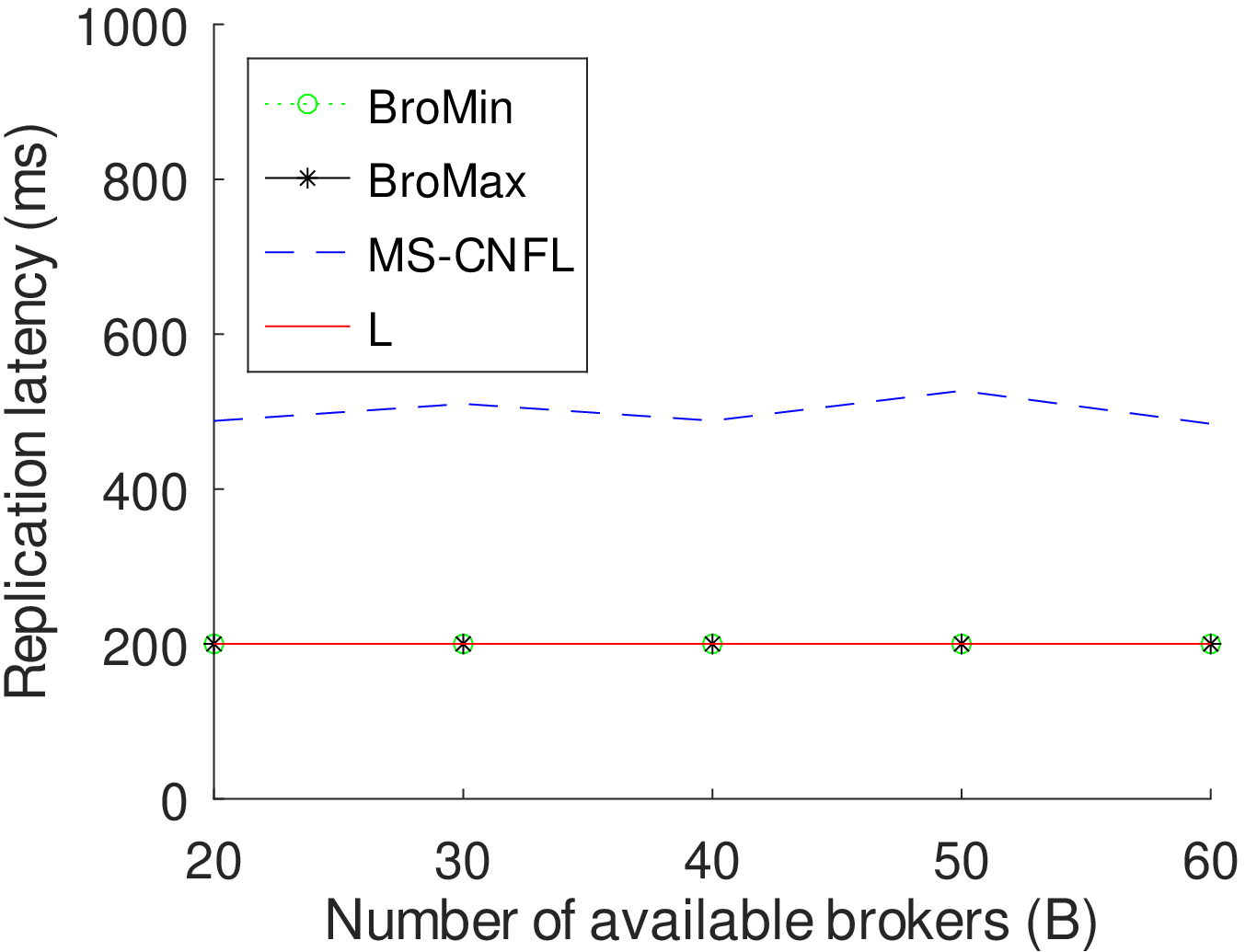}
        \caption{Latency.}
        \label{fig::multibrosim-lat}
    \end{subfigure}
    \begin{subfigure}[b]{0.49\columnwidth}
        \centering
        \includegraphics[width=\columnwidth]{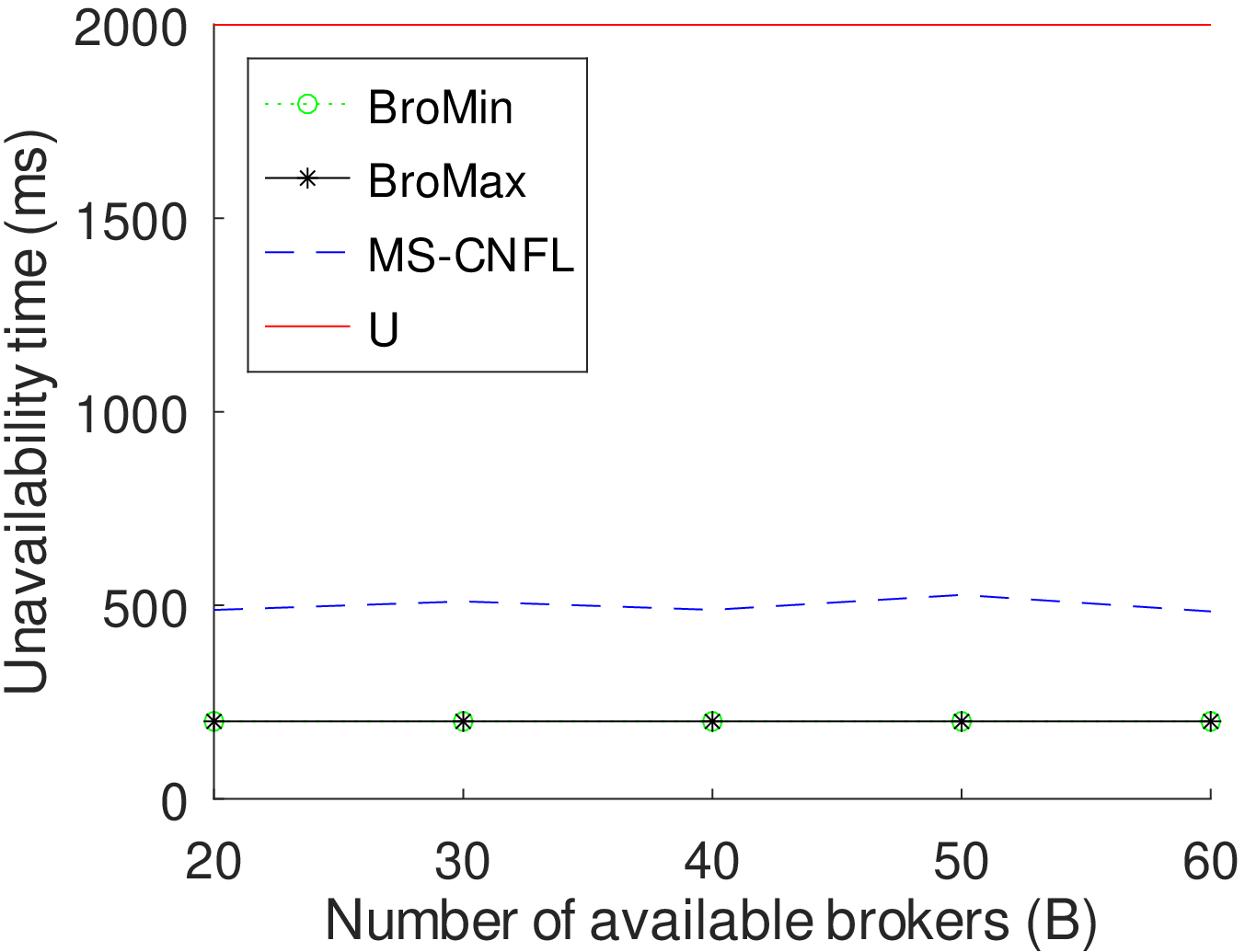}
        \caption{Unavailability.}
        \label{fig::multibrosim-unav}
    \end{subfigure}
    \begin{subfigure}[b]{0.49\columnwidth}
        \centering
        \includegraphics[width=\columnwidth]{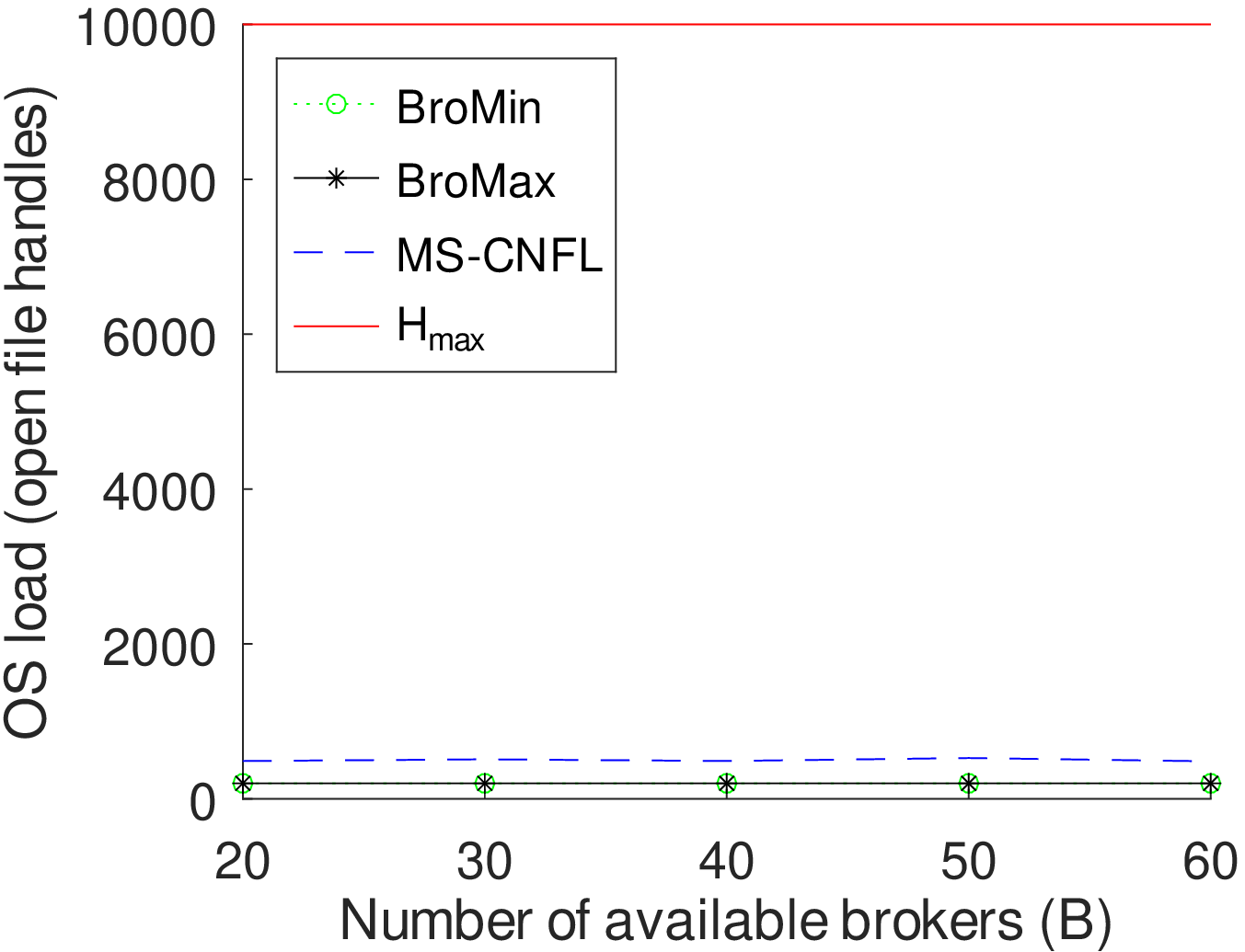}
        \caption{OS load.}
        \label{fig::multibrosim-load}
    \end{subfigure}
    \caption{Performance for variable numbers of available brokers.}\label{fig::multibrosim}
    \vspace{-0.5cm}
\end{figure}

\begin{figure}[t!]
\centering
    \begin{subfigure}[b]{0.49\columnwidth}
    \centering
        \includegraphics[width=\columnwidth]{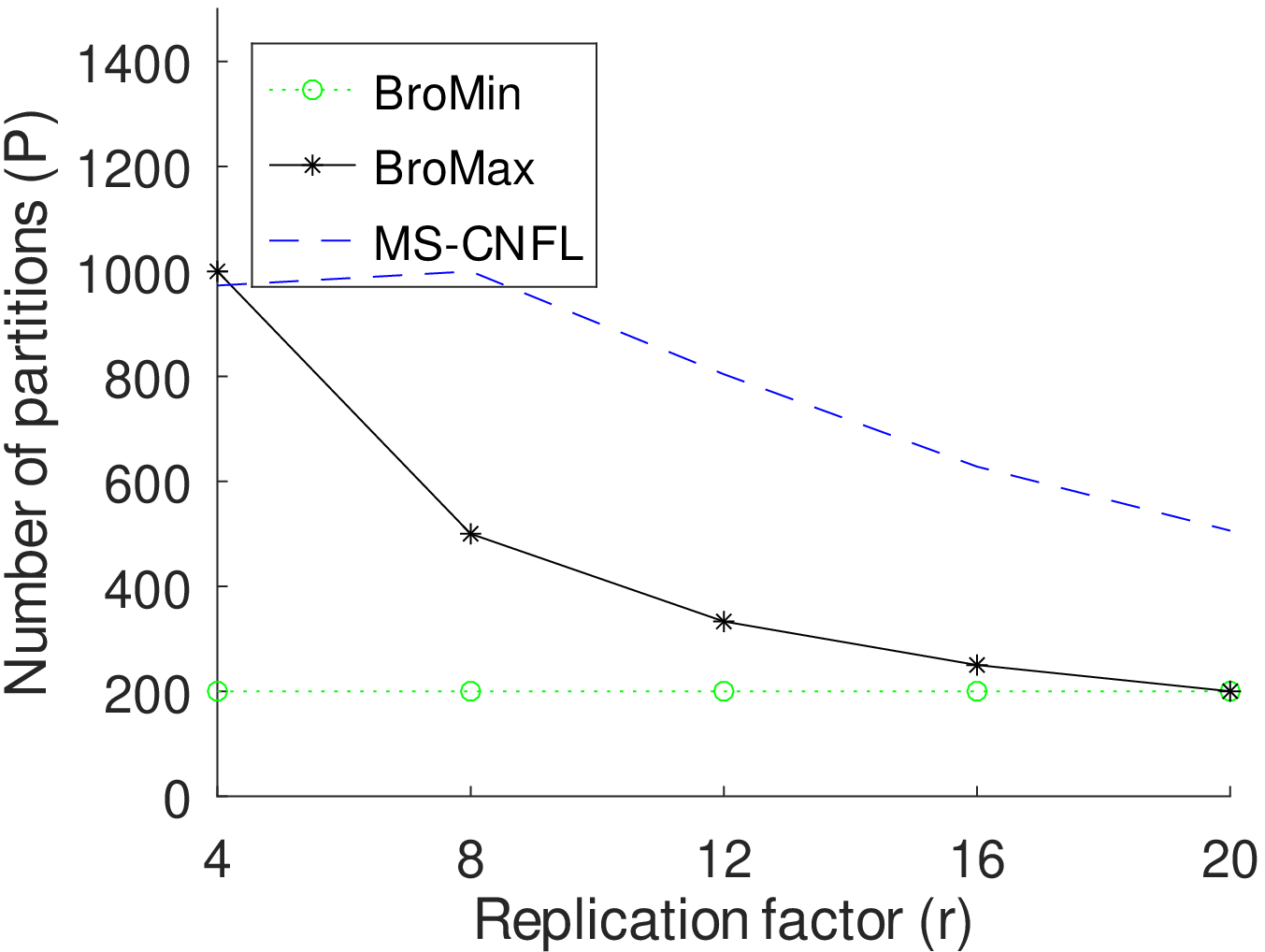}
        \caption{Partitions.}
        \label{fig::multirepsim-part}
    \end{subfigure}
    \begin{subfigure}[b]{0.49\columnwidth}
    \centering
        \includegraphics[width=\columnwidth]{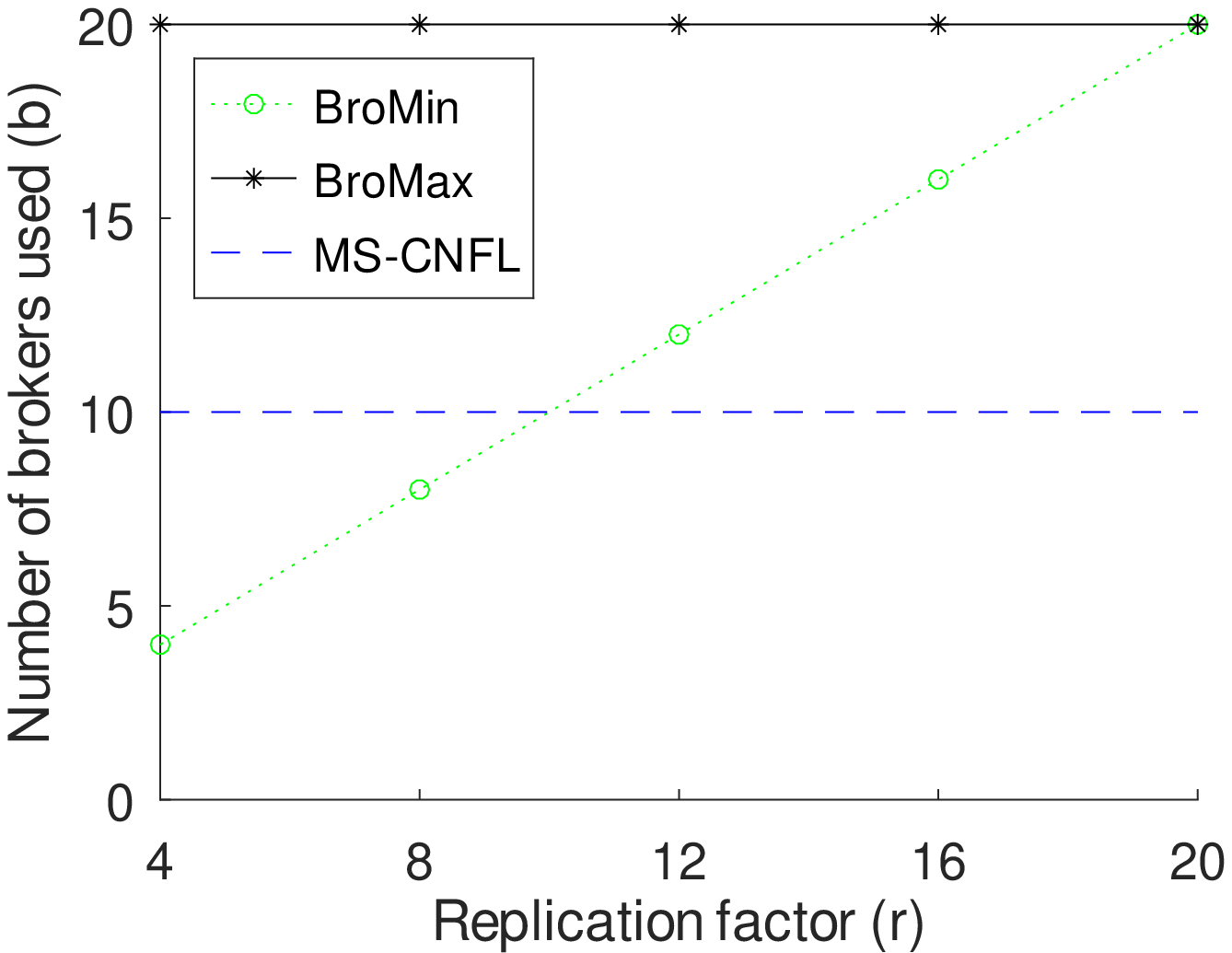}
        \caption{Brokers.}
        \label{fig::multirepsim-brok}
    \end{subfigure}    
    \begin{subfigure}[b]{0.49\columnwidth}
    \centering
        \includegraphics[width=\columnwidth]{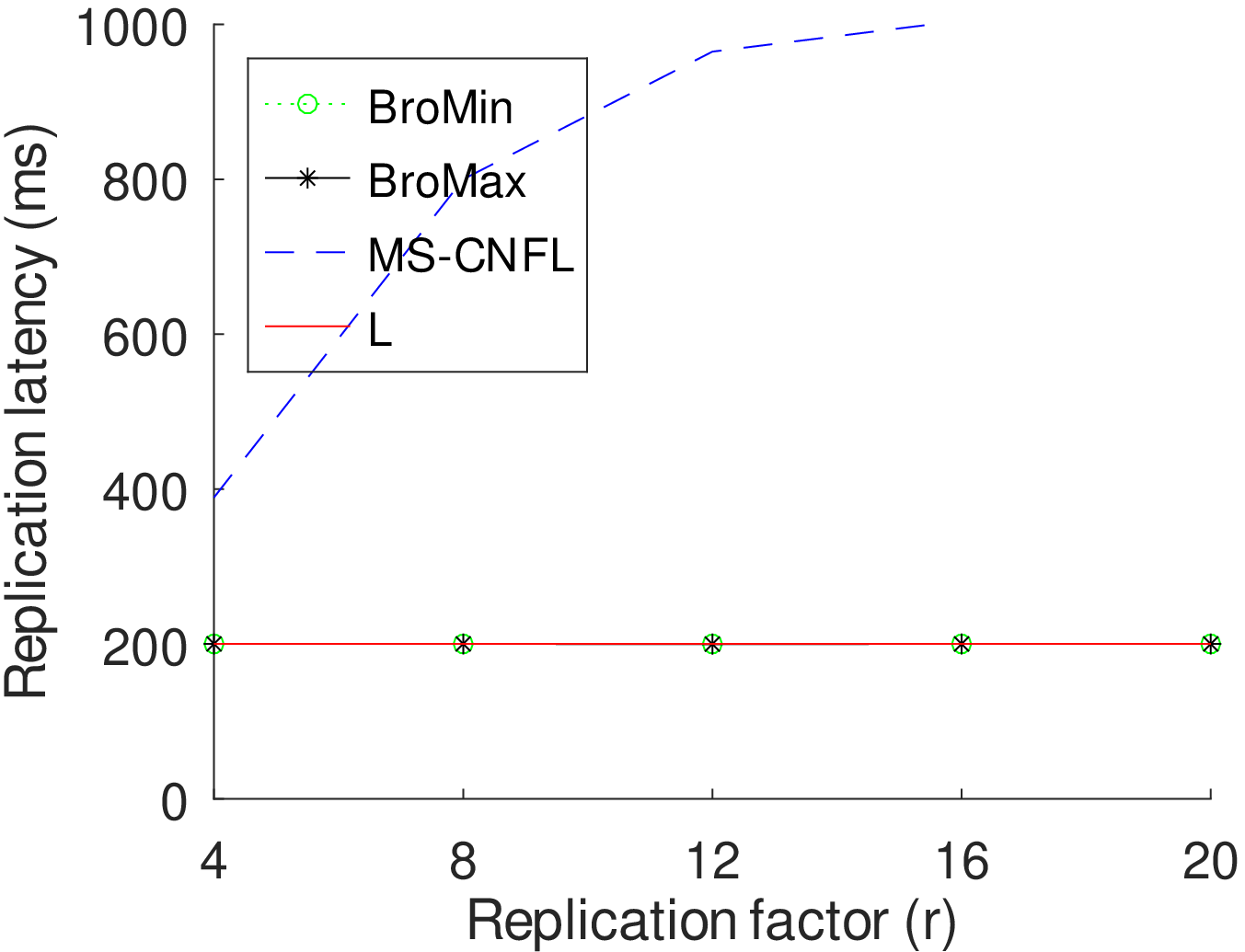}
        \caption{Latency.}
        \label{fig::multirepsim-lat}
    \end{subfigure}
    \begin{subfigure}[b]{0.49\columnwidth}
        \centering
        \includegraphics[width=\columnwidth]{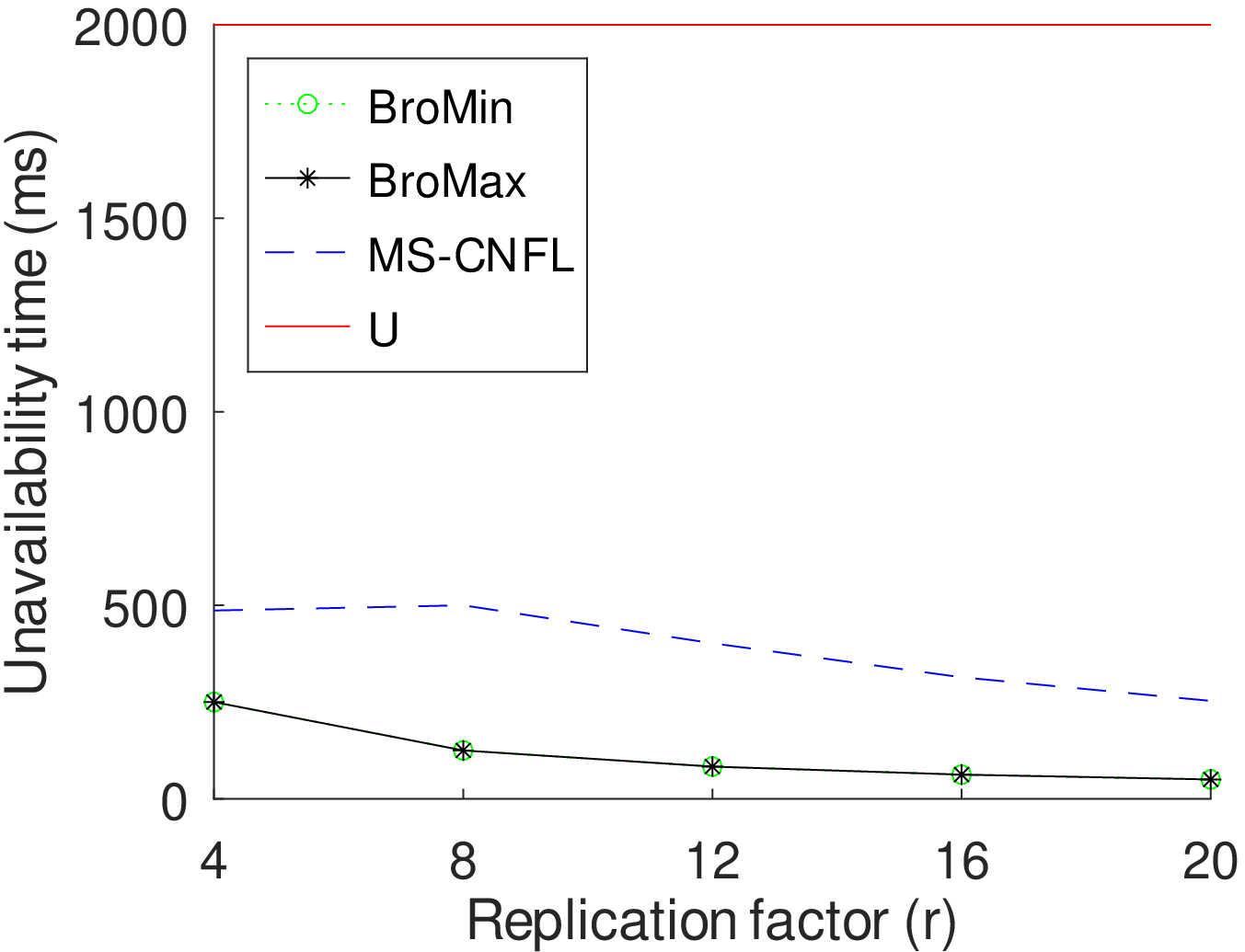}
        \caption{Unavailability.}
        \label{fig::multirepsim-unav}
    \end{subfigure}
    \begin{subfigure}[b]{0.49\columnwidth}
        \centering
        \includegraphics[width=\columnwidth]{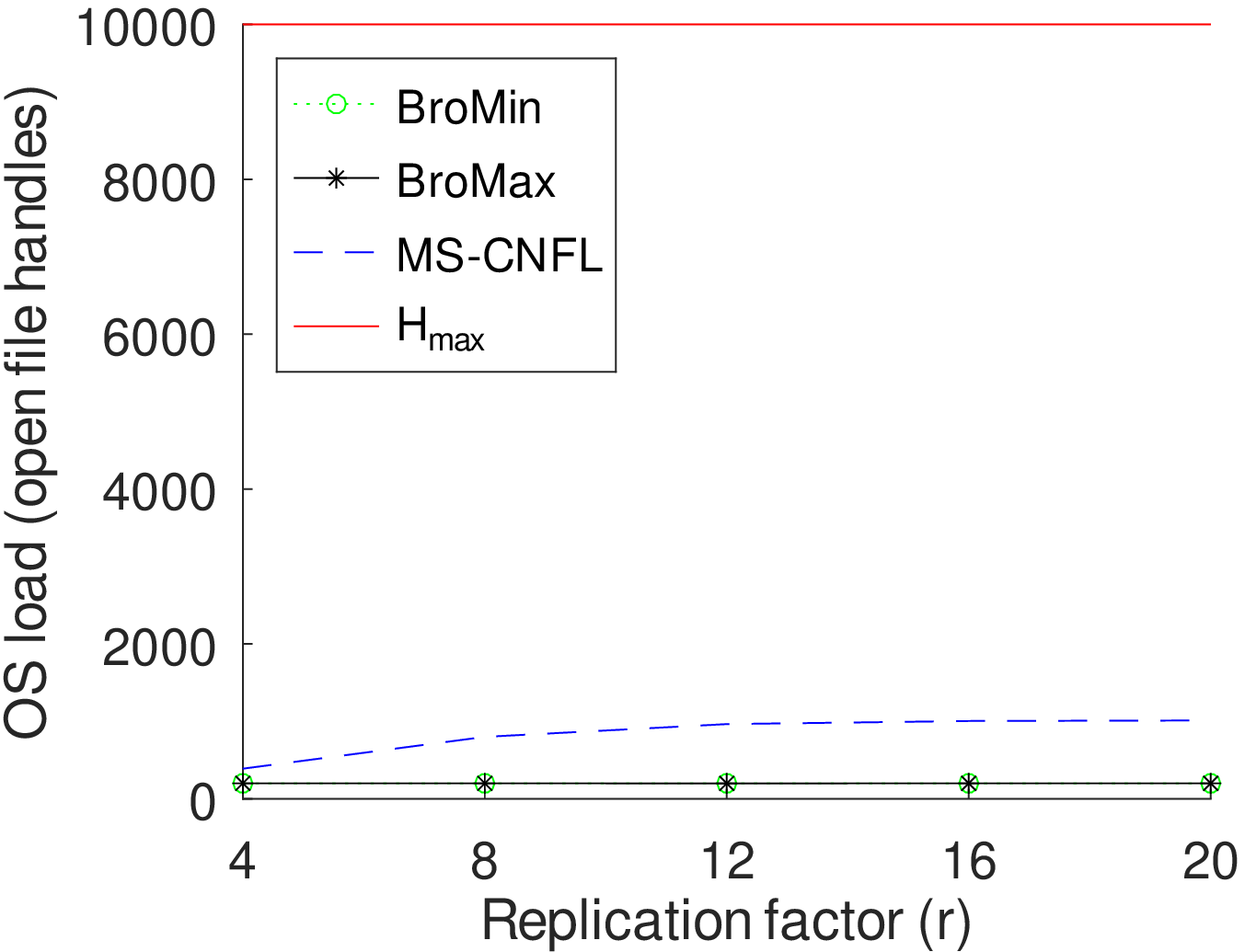}
        \caption{OS load.}
        \label{fig::multirepsim-load}
    \end{subfigure}
    \caption{Performance for variable replication factor values.}\label{fig::multirepsim}
    \vspace{-0.7cm}
\end{figure}

\subsection{Performance metrics and benchmark}

According to \cite{granulate}, there are three main metrics that can reflect the performance of an Apache Kafka cluster:
\begin{enumerate}
    \item System throughput: In our case, this is captured by the ultimate number of partitions selected by each algorithm; as mentioned earlier, the number of partitions is the unit of parallelism in Apache Kafka. In our two algorithms, the desired throughput $T$ is also taken into account as a hard constraint.
    \item Latency: the amount of time that is needed to process data, in the sense of time required for data to be stored or retrieved. The dominant factor in this procedure is the replication time of the data. We capture this metric by appropriately measuring the $l_r$, but also through a hard constraint on $L$.
    \item The numbers or costs of the application's infrastructure: In our case, that would be the number of brokers used in the Apache Kafka cluster. This metric explains the diversity of the two introduced algorithms.
\end{enumerate}

We additionally measure the OS load metric via the open file handles and the unavailability metric via the unavailability time. Given the absence in the literature of a well-defined similar method which targets at efficiently partitioning an Apache Kafka topic, in order to adequately benchmark the performance of our algorithms, we carefully studied the commercial best practices in the related industry. We found out specific Apache Kafka configuration recommendations used by service well known providers. Specifically, Microsoft, is stating as a coarse rule in \cite{microsoft} that it is generally recommended to not exceed $1.000$ partitions per broker (including replicas). Also, Confluent, as a rule of thumb, states in \cite{kafka-topicpartitions}, that if latency is a core objective, it would be probably efficient to limit the number of partitions $P$ per broker $b$ to $100 \cdot B$ ($\cdot r$ if we include the replicas). Therefore, taking into account those two sets of configuration recommendations, we arrive at the following benchmark method, which we call MS-CNFL: $ P = \min \left( P \in_R [1 ... \frac{1.000 \cdot B}{r} ], P \in_R [1...100 \cdot B] \right)$ and $b \in_R [1 ... B]$, where $\in_R$ denotes uniformly random selection. 

\subsection{Performance results}

Figure \ref{fig::multiconsim} displays the four selected metrics for a variable number of consumers. BroMax and MS-CNFL maintain a constant number of partitions, regardless of the number of consumers in the system (Fig.~\ref{fig::multiconsim-part}), with MS-CNFL maintaining a larger number of partitions. On the contrary, BroMin is gradually increasing the number of partitions according to the number of the consumers in the system. A similar behavior is observed in the number of brokers used by each method (Fig.~\ref{fig::multiconsim-brok}). BroMax and MS-CNFL maintain a constant number of brokers, regardless of the number of consumers in the system, and BroMin is gradually increasing the number of brokers according to the number of the available consumers in the system. In this case, we can see that BroMin is choosing to use more brokers than MS-CNFL for more than $400$ consumers in the system. Regarding the hard constraints of latency, unavailability and OS load (Fig.~\ref{fig::multiconsim-lat}, Fig.~\ref{fig::multiconsim-unav} and Fig.~\ref{fig::multiconsim-load} respectively), we can see that while BroMin and BroMax respect both application constraints, MS-CNFL is violating the latency constraint. This performance can be explained by simultaneously observing Fig.~\ref{fig::multiconsim-part} and Fig.~\ref{fig::multiconsim-brok}: MS-CNFL is maintaining more partitions than BroMin and BroMax, but at the same time, tries to distribute them to a non-optimized, insufficient number of brokers. This choice is especially aggravated for more than $400$ consumers in the system, where even BroMin (which tries to minimize the number of brokers required to perform well) uses more brokers than MS-CNFL. But even in the cases where the hard constraints are not violated by MS-CNFL (OS load and unavailability), BroMin and BroMax achieve a better performance with a lower number of partitions.

Figure \ref{fig::multibrosim} displays the four selected metrics for a variable number of available brokers in the cluster. Unlike the previous case of variable consumers, BroMax and MS-CNFL are gradually increasing the number of partitions according to the number of the available brokers in the cluster (Fig.~\ref{fig::multibrosim-part}), with MS-CNFL maintaining a larger number of partitions. On the contrary, BroMin maintains a constant number of partitions regardless of the number of the available brokers in the cluster. A similar behavior is observed in the number of brokers used by each method (Fig.~\ref{fig::multibrosim-brok}). BroMax and MS-CNFL are gradually increasing the number of brokers according to the number of the available brokers in the cluster, and BroMin maintains a constant number of partitions regardless of the number of the available brokers in the cluster. Regarding the hard constraints of latency, unavailability and OS load (Fig.~\ref{fig::multibrosim-lat}, Fig.~\ref{fig::multibrosim-unav} and Fig.~\ref{fig::multibrosim-load} respectively), we can see that, similar to the previous case, while BroMin and BroMax respect both application constraints, MS-CNFL is violating the latency constraint. This performance can be again explained by simultaneously observing Fig.~\ref{fig::multibrosim-part} and Fig.~\ref{fig::multibrosim-brok}: MS-CNFL is maintaining more partitions than BroMin and BroMax, but at the same time, tries to distribute them to a non-optimized, insufficient number of brokers. Even in the cases where the hard constraints are not violated by MS-CNFL (OS load and unavailability), BroMin and BroMax achieve a better performance with a lower number of partitions.

Figure \ref{fig::multirepsim} displays the four selected metrics for a variable replication factor. BroMax and MS-CNFL are gradually decreasing the number of partitions according to the replication factor (Fig.~\ref{fig::multirepsim-part}), with MS-CNFL maintaining a larger number of partitions. On the contrary, BroMin maintains a constant number of partitions regardless of the replication factor. A different behavior is observed in the number of brokers used by each method (Fig.~\ref{fig::multirepsim-brok}). BroMax and MS-CNFL maintain a constant number of brokers, regardless of the replication factor, and BroMin is gradually increasing the number of brokers according to the replication factor. In this case, we can see that BroMin is choosing to use more brokers than MS-CNFL for a replication factor value of more than $10$. Regarding the hard constraints of latency, unavailability and OS load (Fig.~\ref{fig::multirepsim-lat}, Fig.~\ref{fig::multirepsim-unav} and Fig.~\ref{fig::multirepsim-load} respectively), we can see that, in this case, while BroMin and BroMax respect both application constraints, MS-CNFL is not only violating the latency constraint but the violation extent is significantly aggravated for higher replication factor values. This performance can be explained by simultaneously observing Fig.~\ref{fig::multirepsim-part} and Fig.~\ref{fig::multirepsim-brok}: MS-CNFL is maintaining more partitions than BroMin and BroMax, but at the same time, tries to distribute them to a non-optimized, insufficient number of brokers. This choice is especially aggravated for a replication factor of more than $10$, where even BroMin (which tries to minimize the number of brokers required to perform well) uses more brokers than MS-CNFL. But even in the cases where the hard constraints are not violated by MS-CNFL (OS load and unavailability), BroMin and BroMax achieve a better performance with a lower number of partitions.

By combining the findings displayed in each subfigure, we are led to the following conclusions: Both BroMin and BroMax try to use the system resources in a more prudent way than the MS-CNFL configuration recommendations, in the sense that, although MS-CNFL eventually distributes the topic into more partitions, it does not efficiently address the application constraints. In the case of latency constraints, MS-CNFL even violates the given thresholds, especially when the replication factor $r$ becomes higher and higher. On the other hand, BroMin tries to find a delicate balance between the number of partitions and the number of brokers used in the cluster (so as to also enhance efficient resource utilization), and BroMax, using the maximum number of brokers that can be used, tries to maximize the number of partitions by taking into account the application constraints.

\section{Conclusions and future work}
In this paper, after taking into account the Apache Kafka basics, we provided a modeling approach that considers the most important application constraints and requirements. For the first time in the state-of-the-art, given an Apache Kafka topic, we formulated the topic partitioning the problem and we showed its computational intractability, being expressed as an integer program with an objective function of maximizing the topic's partitions. Then, we designed two heuristic algorithms for addressing the problem and we conducted a performance evaluation against Apache Kafka partitioning configuration recommendations provided by Microsoft and Confluent. We demonstrated that both our heuristics use the system resources in a more prudent way than the benchmark method, and address well the constraints of replication latency, resource usage, OS load and unavailability. Our simulation-based study led us to the conclusion that systematic topic partitioning makes sense in Apache Kafka from the point of view of efficiency. As future work, we are planning to perform real-world tests on the Apache Kafka infrastructure that is under development in the EU H2020 MARVEL project.

\balance
\bibliographystyle{IEEEtran}
\bibliography{IEEEfull}

\end{document}